# Observation of superparamagnetism in coexistence with quantum anomalous Hall C=±1 and C=0 Chern states

E. O. Lachman[1], M. Mogi[2], J. Sarkar[1] A. Uri[1], K. Bagani[1], Y. Anahory[1,3], Y. Maysoedov[1], M. E. Huber[4], A. Tsukazaki[5], M. Kawasaki[2,6], Y. Tokura[2,6] and E. Zeldov[1]

[1]Department of Condensed Matter Physics, Weizmann Institute of Science, Rehovot 7610001, Israel

[2]Department of Applied Physics and Quantum Phase Electronics Center (QPEC), University of Tokyo, Tokyo 113-8656, Japan

[3]Racah Institute of Physics, The Hebrew University, Jerusalem 9190401, Israel

[4]Department of Physics, University of Colorado Denver, Denver, Colorado 80217, USA

[5]Institute for Materials Research, Tohoku University, Sendai, 980-8577, Japan

[6]RIKEN Center for Emergent Matter Science (CEMS), Wako, 351-0198, Japan

e-mail: eli.zeldov@weizmann.ac.il, ella.lachman@weizmann.ac.il

**Abstract**

Simultaneous transport and scanning nanoSQUID-on-tip magnetic imaging studies in Cr-(Bi,Sb)$_2$Te$_3$ modulation-doped films reveal the presence of superparamagnetic order within the quantum anomalous Hall regime. In contrast to the expectation that a long-range ferromagnetic order is required for establishing the quantum anomalous Hall state, superparamagnetic dynamics of weakly interacting nanoscale magnetic islands is observed both in the plateau transition regions as well as within the fully quantized $C = \pm 1$ Chern plateaus. Modulation doping of the topological insulator films is found to give rise to significantly larger superparamagnetic islands as compared to uniform magnetic doping, evidently leading to enhanced robustness of the quantum anomalous Hall effect. Nonetheless, even in this more robust quantum state, attaining full quantization of transport coefficients requires magnetic alignment of at least 95% of the superparamagnetic islands. The superparamagnetic order is also found within the incipient $C = 0$ zero Hall plateau, which may host an axion state if the top and bottom magnetic layers are magnetized in opposite directions. In this regime, however, a significantly lower level of island alignment is found in our samples, hindering the formation of the axion state. Comprehension and control of superparamagnetic dynamics is thus a key factor in apprehending the fragility of the quantum anomalous Hall state and in enhancing the endurance of the different quantized states to higher temperatures for utilization of robust topological protection in novel devices.

**Introduction**

The quantum Hall (QH) and the recently demonstrated quantum anomalous Hall (QAH) effects are prime examples of topological states of matter driven by time-reversal symmetry braking (TRSB)[1–3]. In the QH state the TRSB is induced by high magnetic field applied perpendicular to a high-mobility 2D electron gas. The QAH state, in contrast, can be formed in thin films of topological insulators (TI) even in the absence of magnetic field[3–10]. In this case the TRSB is induced by magnetic order attained by doping the TIs with transition metal elements like Cr, V, or Mn[3–8,11,12], which open a mass gap $\Delta$ in the helical Dirac surface states[13–15].

In order to establish a TRSB topological state on a macroscopic scale the magnetic order should be long-ranged and therefore a robust ferromagnetic (FM) state is believed to be a necessary condition for the experimental observation of the QAH effect with dissipationless chiral edge states[2,13,16]. Indeed, a number of studies using global magnetization measurements[14,17,18] and magnetic force microscopy[19,20] have observed FM-like response in magnetically doped TI thin films. Surprisingly, however, recent scanning SQUID magnetic imaging[21] and transport studies[22,23], as well as analysis of Kerr microscopy imaging[24], have suggested the presence of a superparamagnetic (SPM) state in Cr-doped $(Bi,Sb)_2Te_3$ films in which instead of a long-range FM order the magnetism is comprised of single-domain magnetic islands. These islands are of characteristic size of few tens of nanometers and are only weakly coupled to each other[21].

The possible presence of a SPM order has important implications for the properties of the QAH state. The first key consideration is that if the QAH state can indeed be induced by a SPM order, such coexistence can be present only at temperatures well below the SPM blocking temperature, $T \ll T_B$, since at higher temperatures the QAH state will be destroyed by thermally activated magnetization reversals of the SPM islands. At $T \ll T_B$ the overall magnetic response of a SPM is hysteretic similarly to a FM, making it difficult to discern the two types of magnetic order using global magnetization studies. Moreover, the distinction between SPM and FM states may seem to be of minor importance since in both cases a full alignment of the magnetic moments can be attained at low temperatures and at sufficiently high fields above the coercive field $H_c$. This aligned magnetization may then be preserved upon decreasing the field back to zero. However, in addition to causing nontrivial dynamics[22,23,25], the SPM state has two essential repercussions. The first pertains to the existing puzzle of the fragility of the QAH that limits the observation of conductance quantization, $\sigma_{xx} = 0$ and $\sigma_{xy} = e^2/h$, to only very low temperatures that are significantly lower than the magnetic ordering temperature $T_c \cong 20$ K and the size of the magnetically-induced gap in the surface Dirac states[14,15,26], $\Delta \approx 30$ meV. The fact that the SPM islands are only weakly coupled implies reduced magnetism in the regions between the islands which may lead to suppressed $\Delta$ in the inter-island matrix, providing a possible explanation for the observed fragility of the global QAH state.

The second important implication of SPM is related to the recent observation of the $C = 0$ Chern state in which both $\sigma_{xx}$ and $\sigma_{xy}$ vanish[17,27,28] resulting in zero Hall conductivity plateau (ZHP). In contrast to the $C = \pm 1$ QAH case, this intricate insulating state requires the existence of a gap in the TI helical states on the film surfaces but an absence of a global chiral state at the sample edges[29]. One interesting possible mechanism for realization of such a state is the axion insulator[27,30,31], which requires the out-of-plane magnetization to point in opposite directions on the top and bottom surfaces of the film. In this configuration gap opening is induced at the surfaces as well as along the film edges, leading to vanishing $\sigma_{xx}$ and $\sigma_{xy}$. Such a unique magnetic state can be attained in modulation-doped (Bi,Sb)$_2$Te$_3$ films with asymmetric structure resulting in different coercive fields in the top and bottom Cr-doped layers[27]. The recent observation of the enhancement of the ZHP upon increasing the asymmetry of the heterostructures lends strong support to the axion insulator model[27] over other suggested mechanisms, such as the hybridization between the top and bottom surfaces of the TI in presence of disorder[13]. Similarly to the QAH $C = \pm 1$ state, however, the stability of the axion insulator is also crucially dependent on the establishment of a long large magnetic order. Resolving the microscopic magnetic structure in doped TIs is thus paramount for our understanding of the mechanisms leading to the $C = 0$ and $C = \pm 1$ states and for enhancing the robustness of the QAH effect and of the ZHP for device applications.

The degree of the magnetic order in magnetically doped TI films, whether FM, SPM, or disordered, can critically depend on material parameters, doping level, growth conditions, and temperature[11,18,32,33]. In this context, it is important to note that none of the previous studies of doped TI films have directly probed the microscopic magnetic structure within the QAH state[19–21]. The FM state was reported to be present in films that either do not show QAH at all due to non-optimal doping or composition, or at temperatures substantially higher than those required for the formation of the QAH[19,20]. Similarly, the SPM was demonstrated locally in films that do show QAH but at elevated temperatures at which no quantization was present[21,24].

Here we report scanning SQUID-on-tip magnetic imaging of modulation-doped (Bi,Sb)$_2$Te$_3$ films performed concurrently with transport measurements that show full QAH quantization. We observe unambiguous SPM behavior both in the magnetization reversal process in the plateau transition regions near $H_c$ as well as within the macroscopically quantized state. No conventional FM domain wall motion is detected under any conditions. Moreover, our results indicate that SPM order is present also in the incipient $C = 0$ state providing an essential insight into the microscopic nature of the different Chern number topological states.

**Results and discussion**

In this work we studied (Bi,Sb)$_2$Te$_3$ heterostructures with Cr modulation doping which gives rise to a significantly more robust QAH state[34] sustained up to 2 K and in addition renders a clearly developed $C = 0$ state[27]. Figure 1a shows a schematic structure of the film containing two 2 nm thick 12%-Cr doped layers separated by a 3 nm undoped layer. The bottom doped layer is separated from the InP substrate by a 1 nm undoped layer of $(Bi,Sb)_2Te_3$. The heterostructure was coated with 23 nm of $AlO_x$ and 10 nm of Ti/Au top-gate electrode and patterned into a Hall bar structure (Fig. 1b). Imaging of the local out-of-plane magnetic field $B_z(x,y)$ employing a scanning SQUID-on-tip (SOT)[35–37] was carried out simultaneously with transport measurements at $T = 300$ mK. The Indium SOT with effective diameter of 115 nm (Fig. 1c) was scanned at a constant height of about 300 nm above the sample surface at a typical scan rate of 9 μm/sec.

The four-probe transport characteristics of the sample show hysteretic behavior (Fig. 1d) and a clear QAH effect. Full quantization is attained at top-gate bias $V_g = 4$ V when the Fermi energy is tuned into the gap. The apparent coercive field $H_c$, defined by the location of the peaks in $\rho_{xx}$, depends on the sweep rate of the applied field $\mu_0 H$ due to the time relaxation of the magnetization in the film[21]. It reduces from $\mu_0 H_c = 160$ mT for the faster sweep $d\mu_0 H/dt = 95.2$ mT/min used for transport measurements in Fig. 1d to $\mu_0 H_c = 137$ mT at slower effective sweep rate of $d\mu_0 H/dt = 40$ μT/min employed for the simultaneous magnetic imaging and transport (Fig. 1e).

Figure 1e shows an example of scanning SOT magnetic image of $8 \times 4$ μm² acquired at an applied field of $\mu_0 H = 166.4$ mT, slightly above $\mu_0 H_c = 137$ mT. The imaged area includes the edge of the Cr-$(Bi,Sb)_2Te_3$ film (dotted). Since the SOT response is periodic in magnetic field (see Fig. S1) the value of the measured local magnetic field can be determined only up to a constant offset. Here we are interested in studying the variations in local field $B_z(x,y)$ stemming from the spatial fluctuations in the local magnetic structure in the film. We therefore define $B_z = 0$ at ~3 μm from the edge of the sample (at the left end of the imaged area) where the stray field from the magnetized film, which decays as $\sim x/d$, where $x$ is the distance from the edge and $d = 7$ nm is the magnetic thickness of the film, is negligible. The attained $B_z(x,y)$ thus describes the magnetic field above the surface of the sample which arises from the variations in local magnetic structure, while the total magnetic field is given by $\mu_0 H + B_z(x,y)$. At $\mu_0 H = 166.4$ mT $> \mu_0 H_c$ the net average magnetization $M$ of the film is positive resulting in the negative stray field immediately to the left of the sample edge (blue in Fig. 1e). The interior of the sample, however, shows a highly inhomogeneous structure with both positive and negative values of $B_z(x,y)$ reflecting microscopic regions with correspondingly positive and negative out-of-plane magnetization $M_z$.

The magnetism in Cr doped $(Bi,Sb)_2Te_3$ films is strongly anisotropic with an out-of-plane easy magnetization axis[17]. The inhomogeneous magnetic structure in Fig. 1e could therefore arise either from coexistence of FM domains with positive and negative magnetization $M_z$ near $H_c$ or from single-domain SPM islands with random positive and negative moments $m_z$. In order to address this question we acquire a sequence of $B_z(x,y)$ images upon increasing the applied field by small increments of 0.2 mT. In the FM case this procedure should result in expansion of the positive $M_z$ domains into the negatively magnetized regions through domain wall motion. In the case of weakly interacting SPM islands, in contrast, no domain walls should be present and the magnetization process is expected to occur through uncorrelated flipping of the magnetic moments $m_z$ of the different islands determined by their individual coercive fields. Figures 2a,b show an example of two consecutive $B_z(x, y)$ images acquired at $\mu_0 H = 137.4$ and 137.6 mT which appear to be almost identical. In order to determine the small changes we subtract numerically the two images. The resulting $\Delta B_z(x, y)$ image (Fig. 2c) reveals seven distinct circular peaks of positive differential field that are clearly above the background noise level. These peaks can be fitted by point dipoles with magnetic moments of $2m_i$ ranging from $2\times10^5$ to $8\times10^5$ $\mu_B$ as shown in Fig. 2d. This means that seven islands have flipped their magnetic moment $m_z$ from $-m_i$ to $+m_i$ upon increasing the applied field by 0.2 mT. Movie S1 presents a sequence of $B_z(x, y)$ and $\Delta B_z(x, y)$ images upon such incremental sweep of $\mu_0 H$ from 132.6 to 144.2 mT. The same reversal mechanism is revealed at the rest of investigated fields. All the observable magnetization changes thus occur through a random flipping of uncorrelated magnetic islands without any domain wall dynamics, revealing the SPM nature of the magnetic order in the modulation-doped (Bi,Sb)$_2$Te$_3$ films.

The statistical analysis of 785 detected flipping events in four segments of field sweeps are presented in Fig. 2f. An incremental field sweep over 16.2 mT centered around 48.1 mT (blue) revealed 33 events with average magnetic moment of $\bar{m}_i = 1.2\times10^5$ $\mu_B$ and a quite narrow distribution of $m_i$. As the field is increased, the flipping rate grows significantly reaching 485 events in 23.4 mT interval around 141.7 mT (green). In this range, which includes $\mu_0 H_c = 137$ mT, $\bar{m}_i$ increased to $2.1\times10^5$ $\mu_B$ and the distribution becomes significantly broader with few islands reaching $m_i \cong 10^6$ $\mu_B$. Upon further increase of the field to 246 mT (red) the rate drops substantially, the distribution narrows down, and $\bar{m}_i$ decreases to $1.1\times10^5$ $\mu_B$.

It is interesting to compare the SPM magnetization reversal process in our modulation doped samples with the previously reported behavior in uniformly doped Cr-(Bi,Sb)$_2$Te$_3$ films[21]. Figure 2g shows the distribution of the reversal moments in the vicinity of $H_c$ in the two samples. The modulation doped film displays much larger reversal moments reaching $8\times10^5$ $\mu_B$ as compared to about $3\times10^5$ $\mu_B$ for uniform doping. Similarly, $\bar{m}_i = 2.1\times10^5$ $\mu_B$ in the modulation doped film is three times larger than $\bar{m}_i = 7.1\times10^4$ $\mu_B$ in the uniform sample. Taking the similar total Cr doping level, 7 nm film thickness, and assuming 3 $\mu_B$ per Cr atom, the $\bar{m}_i = 7.1\times10^4$ $\mu_B$ in the uniformly doped film translates into SPM island size of $4.9\times10^3$ nm$^2$. In our current modulation-doped sample, the Cr doping level was comparable but the doped layers were only 2 nm thick. The observed $\bar{m}_i = 2.1\times10^5$ $\mu_B$ thus translates into $2.4\times10^4$ nm$^2$ area of a typical SPM island, which is five times larger than in the uniform film. The larger area of the SPM islands implies a substantial reduction of the relative area of the inter-island regions and a more homogeneous magnetization with a likely smaller suppression of Δ in the inter-island matrix, thus providing a possible explanation for the enhanced robustness of the QAH in our modulation doped samples.

By integrating the reversing moments $m_i$ for each field increment we can derive the magnetization curve $M(H) = \sum_i m_i$ (Fig. 2h) along with the simultaneously measured $\rho_{yx}$. Comparison of $M(H)$ and $\rho_{yx}(H)$ curves unveils two significant observations. The first is that in the $C = -1$ plateau region for $\mu_0 H < 80$ mT the magnetization $M$ grows appreciably with $H$ while $\rho_{yx}$ remains quantized. A similar behavior is observed in the $C = +1$ plateau at $H \cong 250$ mT. Figure 2f shows that at these plateau fields substantial island flipping occurs (blue and red distributions) providing a direct observation of SPM dynamics within the QAH $C = \pm 1$ plateau regions. This implies that the global QAH state is preserved in the presence of oppositely magnetized islands as long as their density is low. In Fig. 2h the onset of deviation from quantized $\rho_{yx}$ value occurs at $\mu_0 H \cong 80$ mT (blue arrow) where the relative magnetization of the oppositely magnetized islands reaches about 5%. Above this fraction the edge states surrounding the islands of opposite magnetization apparently start percolating across the sample[13] giving rise to finite $\rho_{xx}$ and a drop in $|\rho_{yx}|$. Our results thus show that the establishment of quantized conductance requires a very large fraction of over 95% of the SPM islands' magnetization to be aligned, significantly higher than required for conventional percolation threshold. The second observation in Fig. 2h is that near the coercive field $\rho_{yx}(H)$ displays a sharper plateau transition than the $M(H)$ curve. This behavior could be ascribed to the enhancement of $\rho_{xx}$ in the $C = 0$ state, similarly to the behavior in the regular anomalous Hall effect where $\rho_{yx} \propto M\rho_{xx}^2$ is commonly found[25].

We now address the $C = 0$ plateau and its magnetic state. Figure 3a shows $\sigma_{xx}$ and $\sigma_{xy}$ attained for the faster field sweep (black) and along several segments of slower sweeps acquired simultaneously with the magnetic imaging (color). At $H_c$ a pronounced suppression of $\sigma_{xx}$ and a plateau-like feature in $\sigma_{xy}$ are visible. Supplementary Fig. S4 shows a more complete set of transport measurements acquired on another sample of the same structure emphasizing the emergence of the ZHP in the $C = 0$ state. A recent study has shown that the $\sigma_{xx} = 0$ and $\sigma_{xy} = 0$ characteristic property of the $C = 0$ state becomes fully developed at low temperatures in modulation doped structures with higher asymmetry between the top and bottom layers[27]. The SOT magnetic imaging in Fig. 2 and Movie S1 acquired concurrently with the transport data clearly show that the magnetic structure of the Cr-(Bi,Sb)$_2$Te$_3$ film in this incipient $C = 0$ state near $\mu_0 H_c = 137$ mT is in SPM state, resulting in highly disordered magnetic configuration and in the absence of domain wall dynamics.

Figures 3b-d compare the magnetic structure in the $C = \pm 1$ and $C = 0$ states. The line cuts in Fig. 3e demonstrate that in the $C = +1$ state most of the SPM islands are magnetized along the positive field direction resulting in an enhanced $B_z(x, y)$ on the inner side of the film and a negative stray $B_z$ outside the edge. Similarly in the $C = -1$

state most of the islands are negatively magnetized resulting in a negative $B_z(x,y)$ on the inner side and a positive stray field outside the film edge. In the $C=0$ state, on the other hand, the densities of the positively and negatively magnetized islands are comparable resulting in zero average $B_z(x,y)$ both inside and outside the film. Moreover, while in the $C=\pm 1$ states in Figs. 3b,d the magnetization is not entirely uniform due to variations in local doping and remaining low density of oppositely oriented islands (see the statistical analysis in Fig. 2f), the degree of the inhomogeneity is relatively small (Fig. 3e). In the $C=0$ state, in contrast, the variations in the local field are substantially larger and $B_z(x,y)$ crosses zero many times, revealing a random distribution of SPM islands with positive and negative values of $m_i$.

In order to realize a robust axion state in TI heterostructures with two doping layers, the magnetic properties of the layers have to meet two requirements: i) The coercive field of one of the layers, $H_c^l$, has to be lower than $H_c^h$ of the second layer, and ii) in the field range $H_c^l < H < H_c^h$, where the two layers have opposite magnetization, the magnetic structure within each of the layers has to be sufficiently homogeneous similarly to the structure attained in the $C=\pm 1$ states with over 95% island alignment. Since in the SPM case the width $\delta H_c$ over which the global magnetization reversal occurs is finite, the second requirement necessitates the separation between the two coercive fields, $\Delta H_c = H_c^h - H_c^l$, to be substantially larger than the width of the magnetization reversal transition of each of the layers, $\Delta H_c \gg \delta H_c$. In such a case, upon sweeping the magnetic field from $C=-1$ state with both layers negatively magnetized (Fig. 3h), the first layer undergoes a transition at $H_c^l$ accompanied by reversal events of the SPM islands. The axion state is established when most of the islands in the first layer have reversed their magnetization, accompanied by a drop in the rate of the magnetization reversals. As the field is further increased upon approaching $H_c^h$, the SPM $m_i$ reversal events in the second layer will set in, leading to the destruction of the axion state. Since our SOT imaging cannot distinguish between the SPM islands in the two layers, we expect to observe two peaks in the rate of $m_i$ reversals at $H_c^l$ and $H_c^h$ and the integrated magnetization curve $M(H) = \sum_i m_i$ should display two step-like features and a plateau at intermediate fields $H_c^l < H < H_c^h$. The transport data showing two peaks in $\sigma_{xx}$ in Fig. 3a is indicative of the positions of the two coercive fields $H_c^l$ and $H_c^h$. In our magnetic imaging data, however, we observe only one broadened range of fields over which the rate of $m_i$ events is large (Fig. S3a) and the corresponding $M(H)$ curve shows a single smoothed step-like behavior with no observable intermediate plateau (Fig. S3b). These results thus indicate that in our modulation doped Cr-(Bi,Sb)$_2$Te$_3$ films the two magnetization reversal transitions overlap substantially, $\delta H_c \sim \Delta H_c$, with no intermediate region in which one layer is close to full positive magnetization and the other close to full negative magnetization (Fig. 3g) thus preventing the establishment of a global axion state. This finding is consistent with the absence of a fully developed ZHP in the transport measurements in Figs. 3a and S4.

**Conclusion**

In conclusion, by performing simultaneous transport and magnetic imaging we reveal that in modulation doped Cr-(Bi,Sb)$_2$Te$_3$ films the SPM state coexists with the QAH state in the $C = \pm 1$ plateaus, in the plateau transition regions, and in the incipient $C = 0$ state. The size of the SPM islands in the modulation doped heterostructures is substantially larger than in comparable homogeneously doped TI films. This points to a more uniform magnetization which apparently lessens the suppression of Δ in the inter-island matrix leading to the observed large enhancement in the temperature robustness of the QAH in modulation doped structures[34]. In order to attain a full conductance quantization over 95% of the SPM islands' magnetization in each layer have to be magnetized along the same direction. In the $C = \pm 1$ state such an alignment can be attained by applying magnetic field that is substantially higher than $H_c$. A similarly high degree of island alignment is required in order to attain a robust axion state. To achieve this, however, the difference in the coercive fields of the two doped layers should be significantly larger than the width of the SPM transition in each of the layers, $\Delta H_c \gg \delta H_c$. The difficulty in attaining this requirement in modulation doped TI films in which the two layers are both Cr doped renders the $C = 0$ state to be significantly more fragile than the counterpart $C = \pm 1$ states. Very recently, two independent studies[38-39] have attained much larger $\Delta H_c$ by doping one layer by Cr and the other layer by V leading to a significantly more robust ZHP providing strong support to this conclusion. Control of the materials properties in terms of doping homogeneity, enlarging the size and the uniformity of the SPM islands, and increasing the difference in the coercive fields of the top and bottom magnetically doped layers thus emerges as the key challenge for attaining robust QAH and axion states and extending their temperature range for novel electronic applications.

**Methods**

**Fabrication of modulation doped Cr-(Bi,Sb)$_2$Te$_3$ heterostructures**

The Cr-(Bi,Sb)$_2$Te$_3$ films were grown by molecular beam epitaxy (MBE) on semi-insulating InP(111) substrates using the same procedures as described in Ref. 27. The 3-nm-thick AlO$_x$ capping layer was deposited by atomic layer deposition (ALD) system immediately after the removal from the MBE chamber. Using standard photolithography and Ar ion milling processes, the films were patterned into the Hall-bar geometry: 600 μm long and 300 μm wide (Fig. 1b). The top 20-nm-thick AlOx dielectric layer and 10-nm-thick Ti/Au electrode were deposited by ALD and electron beam evaporation, respectively.

**SOT imaging**

The magnetic imaging was performed using Indium SOT device[35–37] with an effective diameter of 115 nm (see Supplementary Note 1). The SOT was mounted in a home-built scanning probe microscope employing a series SQUID array amplifier[40] for signal readout. Scanning was performed using Attocube integrated xyz scanner with xy range of 30 µm and z range of 15 µm. Magnetic imaging measurements were performed at 300 mK in Oxford Heliox He$^3$ refrigerator.

The datasets generated during the current study are available from the corresponding authors on reasonable request.

**Acknowledgements**

We thank Karen Michaeli for helpful discussions. This work was supported by the NSF/DMR-BSF Binational Science Foundation – BSF No. 2015653 and NSF No. 1609519, by Rosa and Emilio Segré Research Award, and by JST CREST (No. JPMJCR16F1).

**Contributions**

EOL performed the scanning SOT and transport measurements at 300 mK and constructed the scanning SOT microscope. MM, AT, MK and YT designed, fabricated and characterized the samples. JS performed transport measurements and participated in sample fabrication. AU and KB fabricated the SOTs. YA and YM developed the SOT fabrication technique. YA and EOL developed the data analysis software. MEH developed the SOT readout system. EZ and EOL wrote the manuscript with contributions from other authors.

**Competing Interests**

The authors declare no conflict of interest.

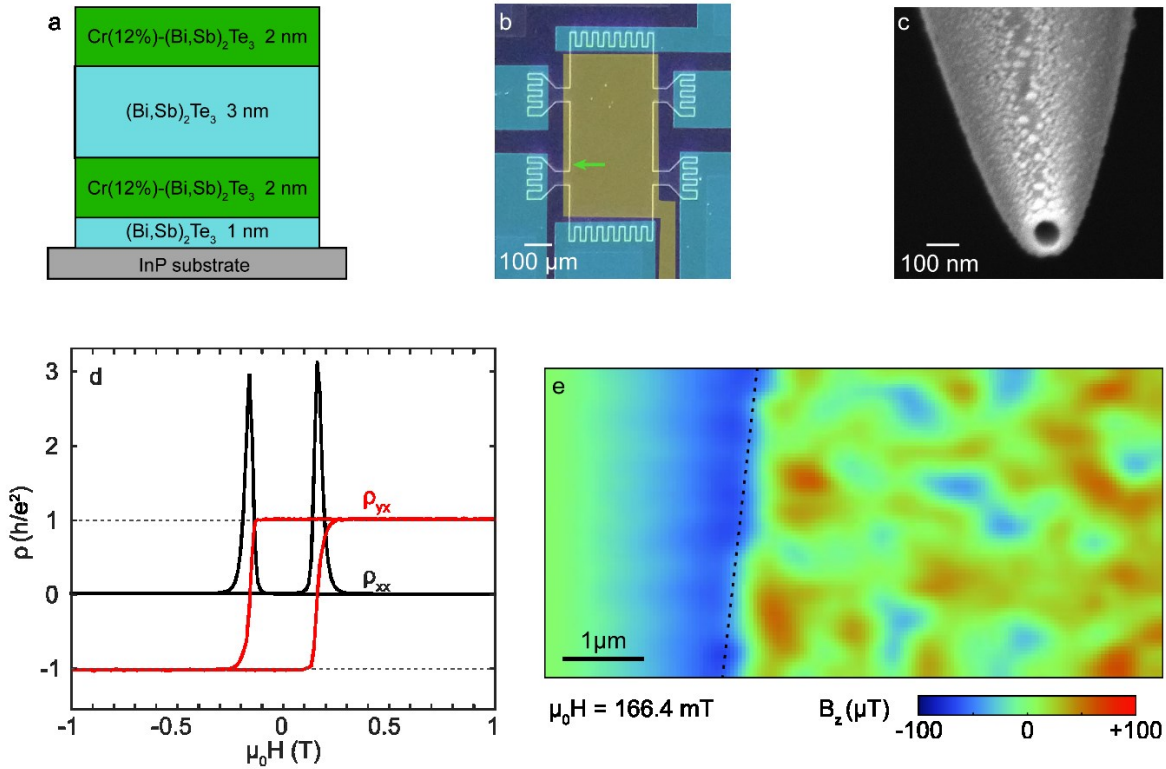

**Figure 1. Characterization of Cr-(Bi,Sb)₂Te₃ modulation-doped heterostructure**. (**a**) Schematic structure of the MBE grown thin film sample. (**b**) False color optical image showing the edges of the Cr-(Bi,Sb)₂Te₃ Hall bar pattern (bright), top gate (yellow), and metallic contacts (blue). The green arrow points to the area at the heterostructure edge imaged in (**e**). (**c**) Scanning Electron Microscope micrograph of the In SOT used for the scanning magnetic imaging. (**d**) Transport measurements of $\rho_{xx}$ (black) and $\rho_{yx}$ (red) showing the QAH state with full quantization at $T = 300$ mK at gate voltage of $V_g = 4$ V. (**e**) Scanning SOT image of the out of plane magnetic field $B_z(x, y)$ above the sample surface at $T = 300$ mK in applied field of $\mu_0 H = 166.4$ mT slightly above the coercive field $\mu_0 H_c = 137$ mT. The dotted line indicates the location of the edge of the patterned heterostructure. Image area 8×4 µm², pixel size 50 nm, sampling time 5.5 ms/pixel.

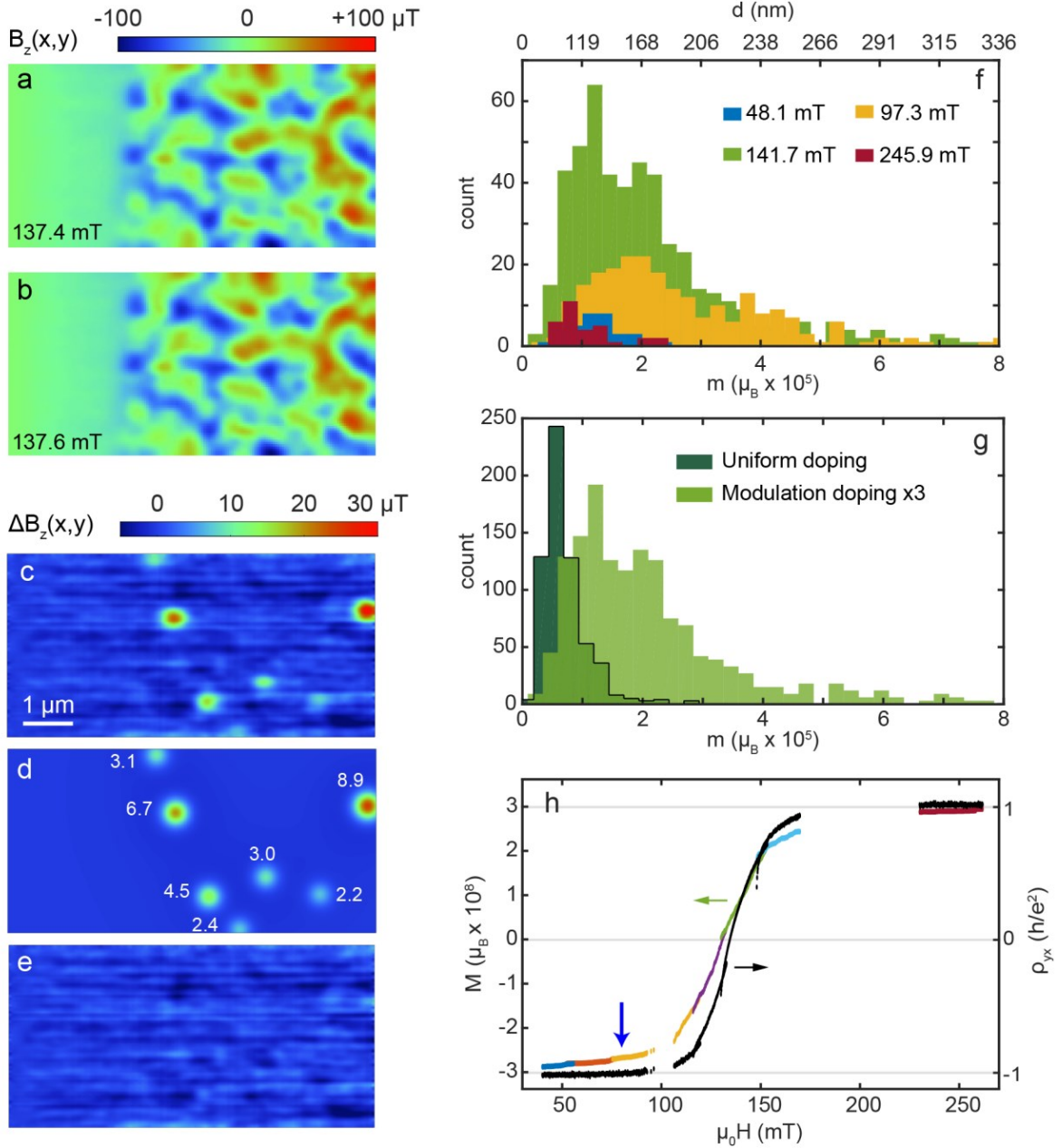

**Figure 2. Superparamagnetism in the modulation-doped Cr-(Bi,Sb)$_2$Te$_3$ heterostructure**. (**a,b**) Scanning SOT $B_z(x,y)$ images 8×4 μm$^2$ at 300 mK at $\mu_0 H = 137.4$ mT (**a**) and 137.6 mT (**b**). (**c**) Differential image $\Delta B_z(x,y)$ obtained by numerically subtracting image (**a**) from image (**b**), showing the magnetization reversal of seven SPM islands (green-red) in response to the 0.2 mT increase in the applied field. (**d**) Best fit numerical simulation of $\Delta B_z(x,y)$ due to seven point dipoles with indicated values of out-of-plane magnetic moment of $2m_i = 2.2 \times 10^5$ μ$_B$ (bottom-right island) to $8.9 \times 10^5$ μ$_B$ (top-right island). (**e**) Residual $\Delta B_z(x,y)$ after subtracting (**d**) from (**c**)

showing the quality of the fit and the noise level of the measurement. (**f**) Histogram of the magnetization reversal moment sizes $m_i$ in four field ranges centered at $\mu_0 H = 48.1$ mT (blue), 97.3 mT (yellow), 141.7 mT (green), and 245.9 mT (red). The top axis shows the corresponding calculated diameter $d$ of the SPM islands. (**g**) Comparison between the moment size histograms of the uniform doped (from Ref.21) and the modulation-doped samples in the vicinity of $H_c$ showing significantly larger SPM islands in the modulation doped heterostructure. (**h**) Cumulative magnetization change $M = \sum_i m_i$ due to island reversals upon sweeping the applied field in seven ranges (color coded, left axis) and the simultaneously acquired $\rho_{yx}$ (black, right axis). The blue arrow at $\mu_0 H \cong 80$ mT indicates the field at which $\rho_{yx}$ starts deviating from quantized value, corresponding to a cumulative change in $M$ of ~5% of the full saturation.

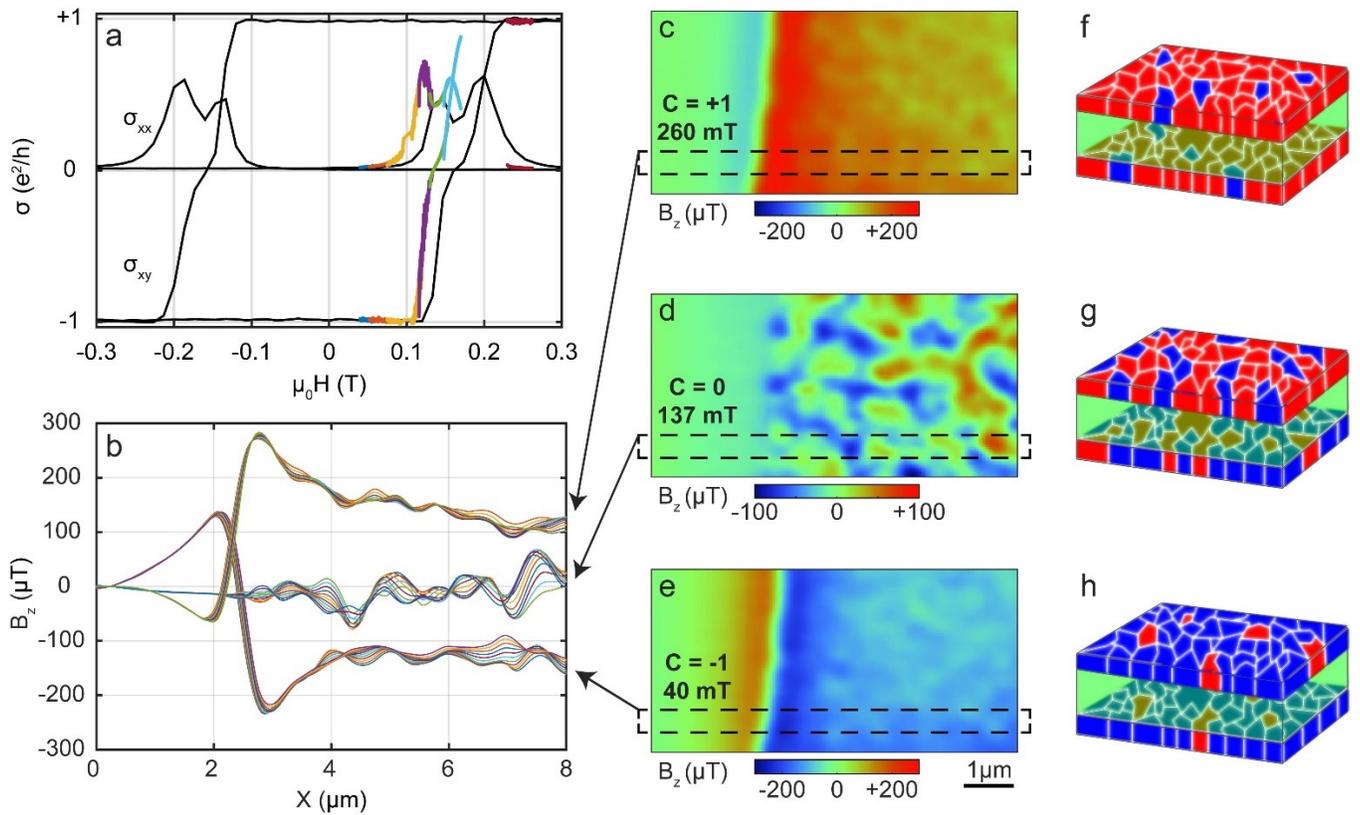

**Figure 3. Magnetic imaging of the $C=0$ and $C=\pm 1$ states.** (**a**) The conductivity coefficients $\sigma_{xx}$ and $\sigma_{xy}$ at 300 mK showing the incipient $C=0$ plateau measured upon a faster field sweep (black) and in a slow sweep acquired simultaneously with the SOT magnetic imaging (color). The slower sweep allows for larger magnetic relaxation and decrease in the apparent coercive field. (**b**) The $B_z(x,y)$ profiles showing the individual scanning SOT line traces within the marked rectangular areas in (**c-e**). (**c-e**) Scanning SOT $B_z(x,y)$ images at $\mu_0 H = 260$ mT (**c**), 137 mT (**d**), and 40 mT (**e**), corresponding to the $C=+1$, $C=0$, and $C=-1$ states respectively. (**f-h**) Schematic configuration of the SPM islands showing high degree of magnetic alignment in the $C=+1$ (**f**) and $C=-1$ (**h**) states. In the incipient $C=0$ axion state (**g**) majority of the islands are positively (negatively) magnetized in the top (bottom) layers but the degree of alignment is insufficient for attaining quantization of transport coefficients.

17